\documentclass[prb,twocolumn,superscriptaddress]{revtex4}

\usepackage{tabularx}
\usepackage{float}
\usepackage{bm}
\usepackage{graphicx}
\usepackage{amsmath}
\usepackage{mathtools}
\usepackage{color}
\usepackage{mathrsfs}
\usepackage{hyperref}
\usepackage{braket}
\usepackage{amssymb}
\usepackage{dsfont}
\usepackage{subfiles}

\begin{document}

\title{Cavity induced chiral edge currents and spontaneous magnetization in two-dimensional electron systems.}

\author{D.D. Sedov}
\affiliation{ITMO University, Kronverkskiy prospekt 49, Saint Petersburg 197101, Russia}
\author{V. Shirobokov}
\affiliation{ITMO University, Kronverkskiy prospekt 49, Saint Petersburg 197101, Russia}
\author{I.V. Iorsh}
\affiliation{ITMO University, Kronverkskiy prospekt 49, Saint Petersburg 197101, Russia}
\author{I.V. Tokatly}
\affiliation{Nano-Bio
  Spectroscopy group and European Theoretical Spectroscopy Facility (ETSF), Departamento de Pol\'imeros y Materiales Avanzados: F\'isica, Qu\'imica y Tecnolog\'ia, Universidad del
  Pa\'is Vasco, Av. Tolosa 72, E-20018 San Sebasti\'an, Spain}
 \affiliation{IKERBASQUE, Basque Foundation for Science, 48009 Bilbao, Spain}
\affiliation{Donostia International Physics Center (DIPC), 20018 Donostia-San Sebasti\'{a}n, Spain}
\affiliation{ITMO University, Kronverkskiy prospekt 49, Saint Petersburg 197101, Russia}

\begin{abstract}
We consider a laterally confined two-dimensional electron gas (2DEG), placed inside a gyrotropic cavity. Splitting of the circularly polarized electromagnetic modes leads to the emergence of the ground state spontaneous magnetization, anomalous Hall effect and chiral edge currents in 2DEG. We examine the dependence of the magnetization and edge current density on the system size for two particular cases of the confining potential: infinite wall and parabolic potentials. We show that paramagnetic and diamagnetic contributions to the edge currents have qualitatively different dependence on the system size. These findings pave the route to the design quantum electrodynamic engineering of the material properties of the mesoscopic electron systems.
\end{abstract}

\maketitle

\section{Introduction}
In recent years, advances in nanofabrication allowed to push the characteristic energies of light-matter interaction in nanosystems embedded in the cavities to the values comparable to cavity photon~\cite{anappara2009signatures,chikkaraddy2016single} energy. This enabled access to the regime of the so-called ultrastrong coupling (USC) between light and matter~\cite{kockum2019ultrastrong}. One of the main consequences of the onset of  ultrastrong coupling is finite occupation number of the cavity photons even in the ground state of the system. This in turn may result in a substantial modification of the material properties when it is embedded in the cavity, which even triggered the emergence of the new field, \textit{Cavity QED materials engineering}~\cite{hubener2020engineering,schlawin2022cavity}. The emergent effects include cavity mediated superconductivity~\cite{thomas2019exploring,curtis2019cavity,sentef2018cavity,schlawin2019cavity,li2020manipulating}, ferroelectric phase transitions~\cite{ashida2020quantum}, topological phase transitions~\cite{PhysRevLett.125.257604, wang2019cavity}, as well as substantial modification of the chemical reactions inside the cavity~\cite{herrera2016cavity,ebbesen2016hybrid,bennett2016novel,martinez2018can,Tokatly2013PRL,ruggenthaler2014quantum,schafer2018ab}.

For most of the systems, the dipole approximation, assuming that the cavity photon field is spatially homogeneous holds since the characteristic wavelength of the cavity photon is typically orders of magnitude larger than the characteristic lengthscale of the material system. At the same time, it has been anticipated, that the phase transitions are forbidden in the cavities with spatially homogeneous modes~\cite{PhysRevB.100.121109, nataf2010no}. Moreover, it has been recently shown that for the macroscopic numbers of two-level systems, the cavity-mediated corrections to macroscopic observables depend only on the effective coupling of a single two-level system to the cavity and thus vanish in the thermodynamic limit~\cite{pilar2020thermodynamics}. At the same time, it is well known, that for a case of a single two-level system placed in a cavity, described by the celebrated Rabi model~\cite{PhysRevLett.107.100401}, ultrastrong light-matter coupling may lead to the substantial modification of the ground state and even to the quantum phase transitions~\cite{PhysRevLett.115.180404}.

Therefore, it would be useful to analyze the dependence of the cavity-mediated corrections to the ground state and various observables on the system size, since currently most of the material systems in the cavity QED experiments belong to the mesoscopic class, comprising large but finite number of particles. 

\begin{figure}[!h]
\includegraphics[width=0.95\columnwidth]{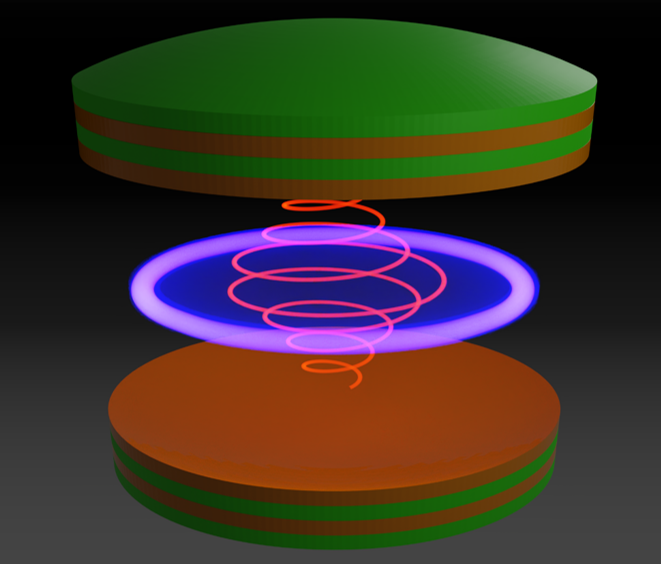}
\caption{\label{fig:f1} 
Geometry of the structure. Two dimensional electron gas is placed inside a Fabry-Perot cavity with ferromagnetic mirrors. Magnetization of the mirrors results in the energy splitting between circularly polarized cavity modes, which, in its turn induces the anomalous magnetization of the 2DEG.}		
\end{figure}

In this work we analyze the cavity-mediated corrections to the ground state and observables for the case of laterally confined two-dimensional electron gas placed in a gyrotropic cavity. We have previously considered this system in the thermodynamic limit and showed that gyrotropy of the cavity leads to the finite Hall conductivity, but the magnetization of the system vanishes in the thermodynamic limit. Here, we derive the explicit dependence of the magnetization on the system size and electronic concentration for the two specific confining potentials: rectangular well and parabolic potential. 

\section{Edge currents in laterally confined two-dimensional electron system}

Previously~\cite{PhysRevB.104.L081408} we have shown that the Hamiltonian effectively describing the electromagnetic field in a gyrotropic cavity maps to the following Hamiltonian of 2D harmonic oscillator placed in the constant magnetic field directed along $z$-axis ($\hbar = 1$),
\begin{align}
    H_{\mathrm{EM}} = \frac{\Omega_0^2}{2} \mathbf{q}^2 + \frac{1}{2} \left[ \boldsymbol{\pi} + \Delta ( \hat{\mathbf{z}} \times \mathbf{q} ) \right]^2,
\end{align}
where $\mathbf{q}$ and $\boldsymbol{\pi}$ are canonically conjugated coordinate and momentum, $[q_i, \pi_j] = i\delta_{ij}$; $\Delta$ is the gyration parameter. Operator of the cavity vector potential is given as $A_{x,y} \propto q_{x,y} \phi(z)$, where $\phi(z)$ is the normalised mode profile which is omitted further by considering the dipole approximation. Two eigenmodes of this Hamiltonian correspond to circulary polarized modes with energy splitting between them equals to $\Delta$.

The Hamiltonian of 2DEG coupled to the cavity electromagnetic (EM) field via constant $\lambda$ reads
\begin{align}
    H_e = \int d^2r \psi^*(\mathbf{r},\tau) \left[ \frac{(\hat{\mathbf{p}} - \lambda \mathbf{q}(\tau))^2}{2m} + U(\mathbf{r}) - \mu \right] \psi(\mathbf{r},\tau)
\end{align}
where $\psi(\mathbf{r},\tau)$ is the imaginary-time electron field operator, $\hat{\mathbf{p}} = -i\nabla$, $\mu$ is chemical potential, and $U(\mathbf{r})$ is a confining potential.

\begin{figure}
    \centering
    \includegraphics[width=\linewidth]{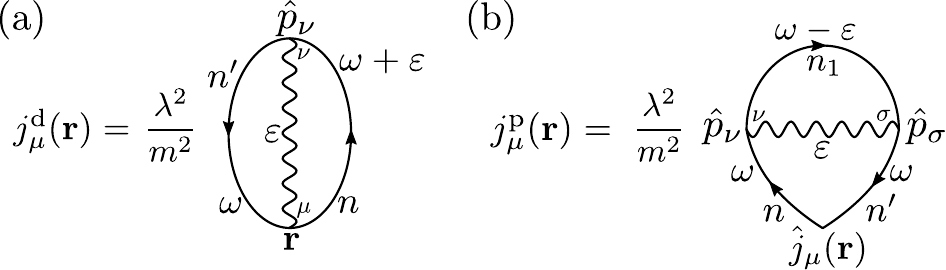}
    \caption{The diagrammatic representation of (a) diamagnetic and (b) paramagnetic parts of the current. Here solid lines correspond to the electron Green's function $G(\omega, n)$, and wavy lines stand for the cavity EM field propagator $D_{\mu\nu}(\varepsilon)$.}
    \label{fig:diagrams}
\end{figure}

We primarily focus on the ground state current induced by the cavity EM field which is given as follows
\begin{align}
    \begin{aligned}
        j_\mu(\mathbf{r}) = &\frac{1}{2m} \Braket{\psi^*(\mathbf{r}) \hat{p}_\mu \psi(\mathbf{r}) - \left( \hat{p}_\mu \psi^*(\mathbf{r})\right) \psi(\mathbf{r})} - \\
        &-\frac{\lambda}{m} \Braket{q_\mu \psi^*(\mathbf{r}) \psi(\mathbf{r})}
    \end{aligned}
\end{align}
where two terms correspond to paramagnetic and diamagnetic contributions respectively. Herein, averaging for an arbitrary operator $A$ is calculated using functional integral approach, $\braket{A} = 1/Z \int D[\psi^*,\psi,\mathbf{q}] A e^{S}$, $Z = \int D[\psi^*,\psi,\mathbf{q}] e^{S}$, with an action $S$ describing evolution in the imaginary time from 0 to $\beta = 1/T$. We calculate the current straightforwardly by expanding the exponent $e^S$ in the powers of $\lambda$. Then the lowest non-vanishing terms corresponding to diamagnetic and paramagnetic contributions shown in Figure~\ref{fig:diagrams}.
It can be readily seen from the Figure~\ref{fig:diagrams}(a) that the diamagnetic contribution is just the product of the photon propagator and the density-current response function
\begin{align}
  j_{\mu}^{\mathrm{d}}(\mathbf{r})=\frac{\lambda^2}{m\beta} \sum_{\varepsilon} D_{\mu\nu}(\varepsilon)\int d\mathbf{r'}\mathbf{\chi}_{n,j_{\nu}}(\mathbf{r},\mathbf{r'},\varepsilon),  
\end{align}
where $D_{\mu\nu}=\langle q_{\mu}(-\varepsilon)q_{\nu}(\varepsilon)\rangle_0$ is the photonic Green's function, and $\chi_{n,j_{\nu}}$ is the corresponding response function. Based on diagram in Fig.~\ref{fig:diagrams}(b) we can similarly express the paramagnetic current as:
\begin{align}
    j_{\mu}^{\mathrm{p}}(\mathbf{r})=\frac{\lambda^2}{\beta} \sum_{\varepsilon} D_{\nu\sigma}(\varepsilon)\int d\mathbf{r'} d \mathbf{r''} \chi_{j_{\mu},j_{\nu},j_{\sigma}}(\mathbf{r},\mathbf{r'},\mathbf{r''},\varepsilon),
\end{align}
where $\chi_{j_{\mu},j_{\nu},j_{\sigma}}$ is the nonlinear response function. This quantity can be directly extracted from the experiment. As we show, in the quantum limit (where the current is induced solely by the vacuum fluctuations of the electromagnetic field) and in the case when the system is  translationally invariant along one of the spatial coordinates, the paramagnetic contribution to the current density vanishes.

The diamagnetic contribution to the current density can be calculated exactly. For that, we can then use the specific sum rule for the density-current response function~\cite{vignale1995sum}: 
\begin{align}
   &\int d\mathbf{r'}\mathbf{\chi}_{n,j_{\nu}}(\mathbf{r},\mathbf{r'},\omega)=\nonumber \\&\frac{1}{m\omega}\left[\nabla_{\mathbf{r}}n_0(\mathbf{r})- \int d\mathbf{r'}\mathbf{\chi}_{n,n}(\mathbf{r},\mathbf{r'},\omega)\nabla_{\mathbf{r'}}V_0(\mathbf{r'})\right], \label{eq:sumrule}
\end{align}
where $n_0$ is the electron density, and $V_0$ is the general potential which may include both the confining potential and possibly disorder potential. If we then calculate the total current $J_{\mu}\sim \int d\mathbf{r}j_{\mu}^{d}(\mathbf{r})$ and consider the single edge confining potential, resulting in translational invariance along one of the coordinates, $y$, we can immediately note that the second term in Eq.~\eqref{eq:sumrule} vanishes due to translational invariance of the density-density response function with respect to $y$. The total edge current can then be expressed as:
\begin{align}
    J_{edge}^y=-\frac{\lambda^2n_0}{m^2\beta}\sum_{\varepsilon}\frac{D_{yx}(\varepsilon)}{\varepsilon}, \label{eq:Jedg}
\end{align}
where $n_0$ is the density of the electron gas in the bulk of the system. We note that this result is universal and holds both for disordered and interacting electronic systems.

The diamagnetic and paramagnetic contributions can be written as  spectral decomposition
\begin{align}
    &\begin{aligned}
        j_\mu^{\mathrm{d}}(\mathbf{r}) = \frac{\lambda^2}{m^2\beta^2}\sum_{\omega,\varepsilon}\sum_{n,n'} & D_{\mu\nu}(\varepsilon) G(\omega+\varepsilon, n) G(\omega, n') \times \\
        &p_{\nu}^{nn'} \psi_{n'}^*(\mathbf{r}) \psi_{n}(\mathbf{r}), \label{eq:diam_c}
    \end{aligned}\\
    &\begin{aligned}
        j_\mu^{\mathrm{p}}(\mathbf{r}) = \frac{\lambda^2}{m^2\beta^2} \sum_{\omega,\varepsilon}\sum_{n,n',n_1} & D_{\nu\sigma}(\varepsilon) G(\omega, n) G(\omega, n') \times\\
        &G(\omega - \varepsilon, n_1) p_{\nu}^{nn_1} p_{\sigma}^{n_1n'} j_{\mu}^{n'n}(\mathbf{r}) \label{eq:param_c}
    \end{aligned}
\end{align}
where $\omega,\varepsilon$ are fermionic and bosonic Matzubara frequencies; index $n$ numerates states of the unperturbed electron system, $G(\omega, n) = - \braket{a_n(\omega) a_n^*(\omega)}_0$ is an electron Green's function of the uncoupled system, where $a_n(\omega)$ is annihilation operator corresponding to the state $\ket{n}$; $\psi_n(\mathbf{r}) = \braket{\mathbf{r} | n}$; $p_{\nu}^{nn'} = \braket{n | \hat{p} | n'}$, $j_{\mu}^{nn'}(\mathbf{r}) = 1/(2m) [\psi_{n'}^*(\mathbf{r})\hat{p}_\mu \psi_{n} (\mathbf{r}) -\psi_{n'}^*(\mathbf{r}) \hat{p}_\mu \psi_{n}(\mathbf{r})]$. 

Summation over fermionic Matzubara frequencies can be performed explicitly, the corresponding result is presented in Appendix A. It is worth to mention mention that for bounded systems, where eigenfunctions can always be chosen purely real, the diagonal part of the photonic propagator does not contribute to the current, and only non-diagonal elements induced by the system's gyrotropy result in non-trivial term.

We now closely examine two particular cases of confining potential: square well and parabolic potentials.

\subsection{Square well potential}

\begin{figure*}[t]
\centering
\center{\includegraphics[width=0.9\linewidth]{
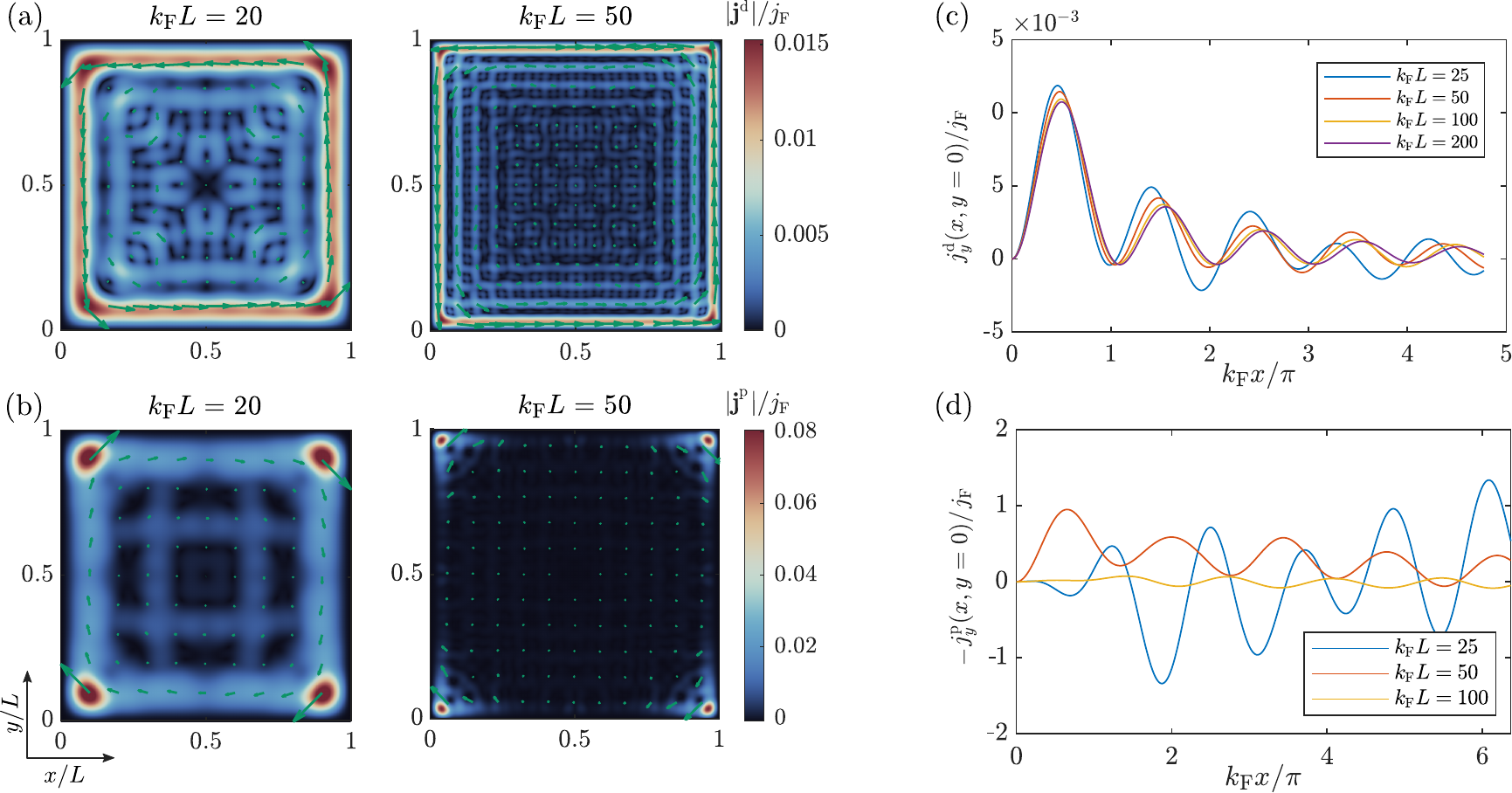}}
\caption{(a) Spatial distribution of the absolute value of diamagnetic current in infinite square well potential for different size parameters $k_FL$. Green arrows shows the direction of current density at given point. (b) Analogous distribution calculated for the paramagnetic current. (c) and (d) present $y$-component of diamagnetic and paramagnetic contributions to the current respectively taken at $y = L/2$ vs $k_{\mathrm{F}}x$. Each dependence is calculated for the following parameters: $\Omega_0 / \mu = 1, \Delta/\mu = 0.01, kT / \mu = 0.01$.}
    \label{fig:Inf_well_current}
\end{figure*}

\begin{figure}
    \includegraphics{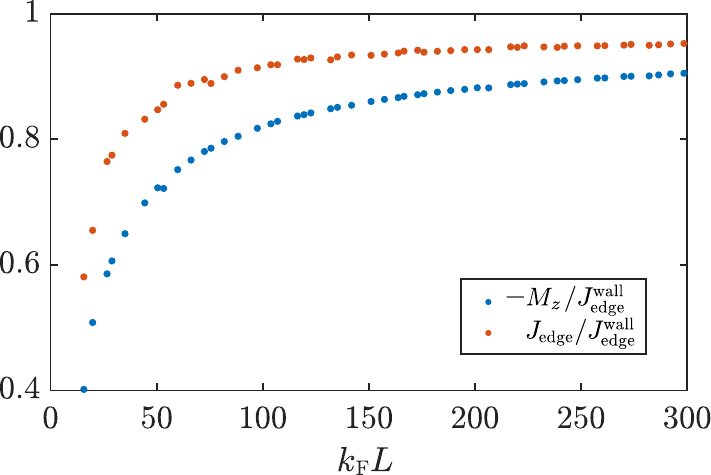}
    \caption{Convergence of edge current and magnetization of the electron system in the infinite well confining potential to the edge current in semi-infinite system. At each plotted point chemical potential $\mu$ and density $n = N/S$ are fixed. The following parameters are used: $\Omega_0 / \mu = 1, \Delta/\mu = 0.01, T/\mu = 0.01$}
    \label{edge_and_magnetization}
\end{figure}

First, we apply the described approach to the case when 2DEG is placed inside a square infinite quantum well with size $L \times L$. An intuitive size-parameter for such system is $k_{\mathrm{F}}L$, where $k_{\mathrm{F}}$ is the length of wavevector on Fermi surface. For large $k_{\mathrm{F}}L$, we expect that effects of the corners to the integral properties and even to the current density almost everywhere can be neglected, the same stands for the the interference between the different edges of the structure. Thus, the general properties in the thermodynamic limit ($k_{\mathrm{F}} L \to \infty$, $k_{\mathrm{F}}$ is fixed) can be understood by considering semi-infinite 2DEG which occupies half-plane $x > 0$. In appendix B we present the derivation of both currents. While paramagnetic density is appeared to be locally zero, diamagnetic current is non-trivial one localized near $x=0$, and it creates the following edge current obtained by the integration over $x$:
\begin{align}
    J_{\mathrm{edge}}^{\mathrm{wall}} = - \frac{\lambda^2 n_0 \Delta}{2 \Omega_0^2 \tilde{\Omega}_0 m^2} \left[1 + n_{\mathrm{B}}^{+} + \frac{\tilde{\Omega}_0}{\Delta} n_{\mathrm{B}}^{-} \right], \label{eq:one_wall_edge}
\end{align}
where $\tilde{\Omega}_0 = \sqrt{\Omega_0^2 + \Delta^2}$, $n_\mathrm{B}^\pm = n_{\mathrm{B}}(\Omega_-) \pm n_{\mathrm{B}}(\Omega_+)$, $\Omega_{\pm} = \tilde{\Omega}_0 \pm \Delta$, $n_0$ is the density of electrons in the system, and $n_{\mathrm{B}}$ is Bose-Einstein distribution. We note that this result is very similar to the one obtained for the case of the edge dc current in the 2DEG under classical circularly polarized optical pump in the case of weak disorder potential~\cite{durnev2021rectification}

In the fig.~\ref{fig:Inf_well_current} (a) and (b) we present spatial distribution of the absolute values of both currents, their vector plots are depicted with green arrows. We normalize our results on $j_{\mathrm{F}} = n_0 v_\mathrm{F} (\Delta / \mu) (\lambda^2 / m \mu^2)$, where $v_{\mathrm{F}} = \sqrt{2\mu / m}$ -- electron's speed on the Fermi surface. Numerical calculations show that currents have significantly different bahaviours. Diamagnetic current relatively far away from corners tends to some distribution which can be clearly see in the fig.~\ref{fig:Inf_well_current} (c), while paramagnetic current density has non-zero distribution only near corners.  

We also present the convergence of the edge current and magnetization of the system to $J_{\mathrm{edge}}^{\mathrm{wall}}$ with respect to $k_{\mathrm{F}} L$ in the figure~\ref{edge_and_magnetization}. It proves our considerations that in thermodynamic limit the system is mainly defined by the properties of semi-infinite 2DEG.

\subsection{Harmonic Oscillator}

\begin{figure*}[t!]
    \includegraphics[width=0.9\linewidth]{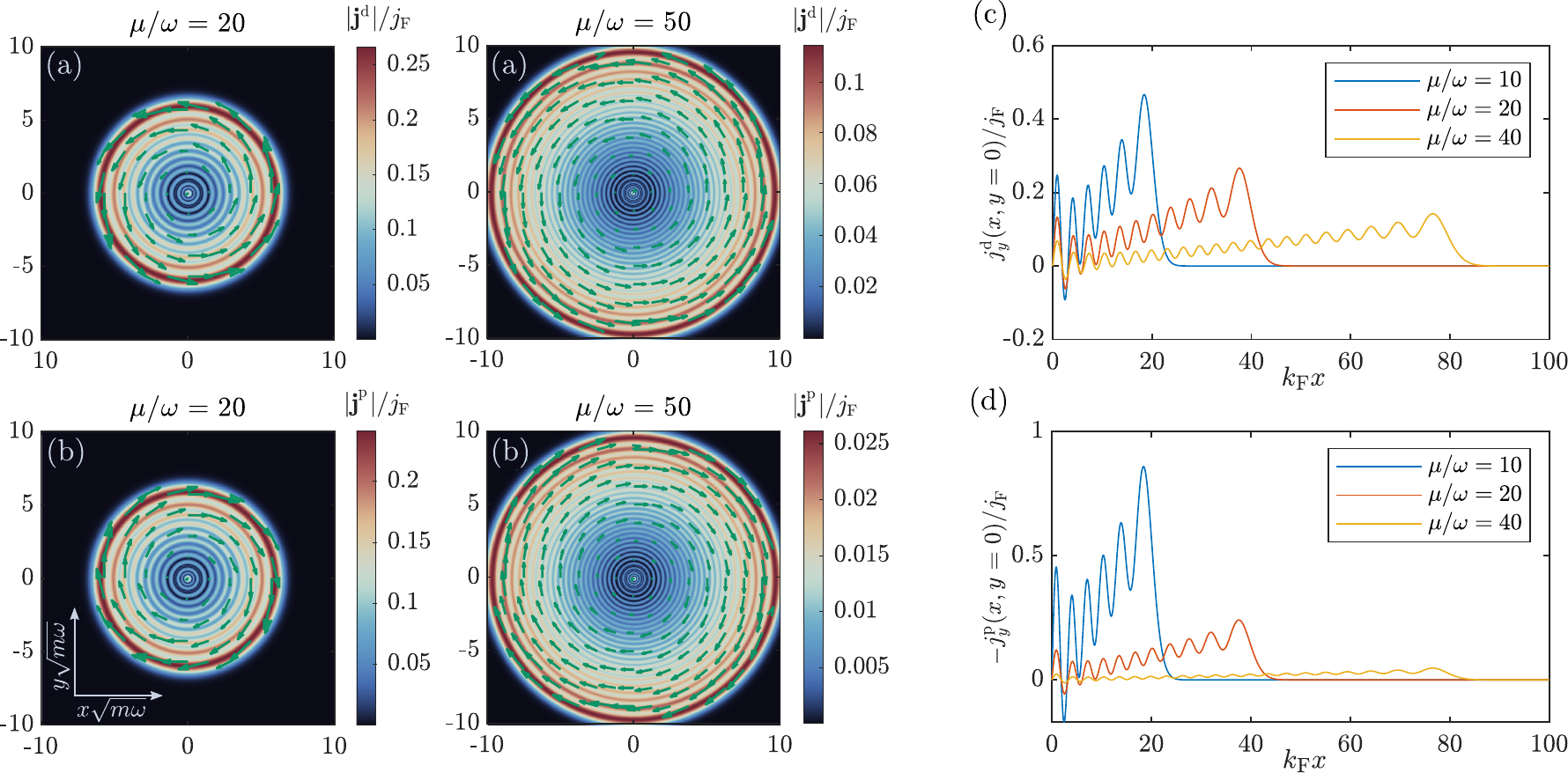}
    \caption{(a) Spatial distribution of the diamagnetic part of the current in the parabolic potential for two values of $\mu / \omega$. (b) Analogous distribution calculated for the paramagnetic current. (c) and (d) present radial distribution of the diamagnetic and paramagnetic contributions to the current respectively. Each dependence is calculated for the following parameters: $\Omega_0 / \mu = 1, \Delta/\mu = 0.01, kT / \mu = 0.01$.}
    \label{fig:harmonic_osc_current}
\end{figure*}

Let us consider parabolic confining potential $U(\mathbf{r}) = m \omega^2 \mathbf{r}^2 / 2$ -- another relatively simple system which allows to obtain some analytical results not even in the thermodynamic limit.

\begin{align}
     &\begin{aligned}
        j^{\mathrm{d}}_y(\mathbf{r}) = \frac{4\lambda^2}{m^2\beta} \sqrt{\frac{m\omega}{2}} \sum_{n}& [f_{n+1} - f_n] \sqrt{n_x + 1} \psi_{n_x+1}(x) \times\\
         & \psi_{n_x}(x) |\psi_{n_y}(y)|^2 \sum_{\varepsilon} \frac{\varepsilon D_{yx}(\varepsilon)}{\varepsilon^2 + \omega^2},
     \end{aligned}\\
     &\begin{aligned}
         j_y^{\mathrm{p}}(\mathbf{r}) = \frac{4\lambda^2 \omega^2}{m\beta} \sum_n & \psi_{n_x+1}(x) \psi_{n_x}(x) \times \\
         &\mathrm{Im} \left[ j_y^{n_y+1, n_y}(y) \right] \sqrt{(n_x + 1)(n_y + 1)} \times\\
         &\left[ f_{n+2} - 2f_{n+1} + f_n \right] \sum_\varepsilon \frac{\varepsilon D_{yx}(\varepsilon)}{(\varepsilon^2 + \omega^2)^2},
     \end{aligned}
\end{align}
where $\psi_{n_{x}}(x)$ are eigenfunctions of one dimensional harmonic oscillator, $f_n = f(\omega n)$ is Fermi-Dirac distribution. From the obtained expression one clearly sees that in the low temperature limit, $T/\mu \to 0$, currents are determined only by the electrons on the Fermi surface.

We present the distributions of currents in the figure~\ref{fig:harmonic_osc_current}. For both contributions, the radius of localization, where currents are not exponentially suppressed, grows as $O(1/\omega)$. It can be understood from the quasiclassical approach in which this radius is given by $\sqrt{2\mu / (m\omega^2)} $. In the same time, the amplitudes of two parts scale differently with respect to $\omega$, and consequently, result in  which can be seen in the spontaneous magnetic moments
\begin{align}
    m_z^{\mathrm{d}} &= \left(\left\lfloor \frac{\mu}{\omega} \right\rfloor^2 + \left\lfloor \frac{\mu}{\omega} \right\rfloor \right) \frac{\lambda^2}{2m^2\beta} \sum_{\varepsilon} \frac{\varepsilon D_{yx}(\varepsilon)}{\varepsilon^2 + \omega^2},\\
    m_z^{\mathrm{p}} &= -\omega^2 \left( \left\lfloor \frac{\mu}{\omega} \right\rfloor^2 + 5 \left\lfloor \frac{\mu}{\omega} \right\rfloor + 6 \right) \frac{\lambda^2}{3m^2\beta} \sum_{\varepsilon} \frac{D_{yx}(\varepsilon)}{(\varepsilon^2 + \omega^2)^2}.
\end{align}
Explicit result obtained after analytical summation over Matzubara frequency is cumbersome and presented in Appendix C. These expressions hold for an arbitrary relation $\mu/\omega$ and require only small temperatures relative to the chemical potential. Since number of occupied states in the system is equal to $1/2 \lfloor \mu / \omega \rfloor ( \lfloor \mu / \omega \rfloor + 1)$, we obtain non-zero magnetization in the limit of vanishing oscillator frequency, $\omega/\mu \to 0$, only for the diamagnetic contribution. It coincides with the results we have found for the infinite well potential. It is interesting that system with completely different potential inherits this behaviour.

\section{Vacuum anomalous Hall effect in the presence of disorder and cavity losses}
In this section we derive the Hall conductivity of the system from the equations of motions for the observables taking into account disorder in electron gas and cavity losses. First of all, we note that the current is connected to the total electric field $\mathbf{E}$ via the Drude formula $\mathbf{j}=\sigma_D \mathbf{E}$, where $\sigma_D=(ne^2/m)(1-i\omega\tau)^{-1}$, $\tau$ is the momentum relaxation time. The total field comprises the external field $\mathbf{E}_0$ and the cavity field $\mathbf{E}_{cav}$. To find the cavity field we can write the Hamiltonian of the system
\begin{align}
    H=\frac{1}{2m}(\mathbf{p}-\lambda\mathbf{q}-e\mathbf{A}_0)^2+\frac{1}{2}(\boldsymbol{\pi}+\Delta \hat{z}\times \mathbf{q})^2+\frac{1}{2}\Omega_0^2q^2,
\end{align}
where $\mathbf{r},\mathbf{p}$ and $\mathbf{q},\boldsymbol{\pi}$ are canonical coordinate and momentum for electron and cavity photon respectively. We can immediately see that the cavity electric field $\mathbf{E}_{cav}=\lambda\dot{\mathbf{q}}/e$. The equation of motion for the photon coordinate $q$ is given by
\begin{align}
    &\ddot{\mathbf{q}}-2\Delta \hat{z}\times\dot{\mathbf{q}}+\Omega_0^2q+\gamma \dot{\mathbf{q}}=\lambda\mathbf{v}\label{class:q},
\end{align}
where $\mathbf{v}=\mathbf{j}/(ne)$ is the electorn velocity, and $\gamma$ is the cavity losses rate. We assume that cavity losses very weakly depend on frequency. Since the equation is linear, we can take the Fourier transform and arrive to a linear system of equations. From Eq.~\eqref{class:q} we find the expression for the cavity field:
\begin{align}
    \mathbf{E}_{cav}=-i\omega \lambda^2/(ne^2) \hat{D}\mathbf{j},
\end{align}
where $\hat{D}$ is the cavity photon propagator,
$[\hat{D}^{-1}]_{xx}=[\hat{D}^{-1}]_{yy}=\Omega_0^2-\omega(\omega+i\gamma)$ and $[\hat{D}^{-1}]_{xy}=-[\hat{D}^{-1}]_{yx}=-2i\omega\Delta.$
The expression for the current then reads
\begin{align}
    \mathbf{j}=\left[\hat{I}+\frac{\omega\tau}{i+\omega\tau}\frac{\lambda^2}{m}\hat{D}\right]^{-1}\sigma_D\mathbf{E}_0,\label{eq:classical}
\end{align}
where $\hat{I}$ is the unity matrix, and $\sigma_D=(ne^2\tau/m)(1-i\omega\tau)^{-1}$ is the Drude conductivity. From Eq.~\eqref{eq:classical} one immediately can see that that the contribution to the DC current from the coupling to the cavity photon vanishes for any finite $\tau$. This effect is similar to vanishing of the spin Hall current at any finite disorder~\cite{mishchenko2004spin,ol2005spin,raimondi2005spin}. We note, that while the Hall conductivity vanishes at any finite disorder, the stationary current present in the absence of the external field, is immune to disorder as shown in Section II.

\section{conclusion}
We have considered the generation of the chiral edge currents and spontaneous magnetization in the mesoscopic system comprising a laterally confined 2D electron gas placed inside a gyrotropic cavity. It has been shown, that the diamagnetic and paramagnetic contributions to the edge current density have qualitatively different asymptotic behaviour when approaching the thermodynamics limit: while paramagnetic current density vanishes locally, the diamagnetic contribution approaches finite value which results in the finite magnetization in the thermodynamic limit. We have also shown using semiclassical equations that the arbitrarily small disorder in 2DEG destroys the DC Hall conductivity, while leaving the ac Hall conductivity finite. These results suggest the cavity engineering of the
material properties can serve as a powerful tool for controlling the transport properties in the mesoscopic systems.

\begin{widetext}

\appendix

\section{Explicit expressions for currents}

The explicit expression for the electron Green's function $G(\omega, n) = (i\omega + \mu - E_n)^{-1}$ and summation over fermionic Matzubara frequencies $\omega = 2\pi s T, s \in \mathds{Z}$, allows to rewrite~\eqref{eq:diam_c} in the following way
\begin{align}
    &j_{\mu}^{\mathrm{d}}(\mathbf{r}) = \frac{\lambda^2}{m^2\beta} \sum_{\nu,\varepsilon} \sum_{n,n'} \frac{D_{\mu\nu}(\varepsilon)}{i \varepsilon + E_{n'} - E_n} [f(E_n') - f(E_n)]  \psi_{n'}^*(\mathbf{r}) \psi_{n}(\mathbf{r}) p_{\nu}^{nn'}, \label{eq:paramagnetic_simplified}
\end{align}
where $f(E_n)$ is Fermi-Dirac distribution. Using the properties of the photonic propagator $D_{xx}(\varepsilon) = D_{yy}(\varepsilon), D_{yx}(\varepsilon) = -D_{yx}(-\varepsilon) = - D_{xy}(\varepsilon)$, we can divide the diagonal and non-diagonal contributions of the cavity EM field,   
\begin{align}
    \begin{aligned}
        j_{\mu}^{\mathrm{d}}(\mathbf{r}) & = \frac{\lambda^2}{m^2\beta} \sum_{\varepsilon, n,n'} \frac{D_{\mu\mu}(\varepsilon)}{\varepsilon^2 + (E_{n'} - E_n)^2} [f(E_{n'}) - f(E_n)][E_{n'} - E_n] \mathrm{Re}\left[ \psi_{n'}^*(\mathbf{r}) \psi_{n}(\mathbf{r}) p_{\mu}^{nn'} \right] + \\
        &+ \frac{\lambda^2}{m^2\beta} \sum_{\varepsilon, n,n'} \frac{ \varepsilon \left. D_{\mu\nu}(\varepsilon) \right|_{\nu \neq \mu}}{\varepsilon^2 + (E_{n'} - E_n)^2} [f(E_{n'}) - f(E_n)] \mathrm{Im}\left[ \psi_{n'}^*(\mathbf{r}) \psi_{n}(\mathbf{r}) p_{\nu}^{nn'} \right].
    \end{aligned}
\end{align}
Since for a bounded system eigenfunction can always be made purely real, the diagonal. With analogous considerations one can obtain the following expression for the paramagnetic part of the current
\begin{align}
    \begin{aligned}
        j_\mu^{\mathrm{p}}(\mathbf{r}) = \frac{\lambda^2}{\beta m^2}\sum_{\nu\neq\sigma, \varepsilon}\sum_{n, n', n_1} \frac{D_{\nu\sigma}(\varepsilon) \varepsilon}{E_n - E_n'} & \left[ \frac{f(E_n)}{(E_n - E_{n_1})^2 + \varepsilon^2} - \frac{f(E_{n'})}{(E_{n'} - E_{n_1})^2 + \varepsilon^2} - \right. \\
        & \left. \frac{f(E_{n_1}) (E_n - E_{n_1})(E_n + E_{n'} - 2E_{n_1})}{\left[ (E_n - E_{n_1})^2 + \varepsilon^2 \right]\left[ (E_{n'} - E_{n_1})^2 + \varepsilon^2 \right]} \right] \mathrm{Im} \left[  p_{\nu}^{nn_1} p_{\sigma}^{n_1n'} j_{\mu}^{n'n}(\mathbf{r})\right]. \label{eq:paramagnetic_current}
    \end{aligned}
\end{align}
If a bounded system can be factorized with respect to coordinates, this expression can be simplified
\begin{align}
    j_y^{\mathrm{p}}(\mathbf{r}) = \frac{2\lambda^2}{\beta m^2}\sum_{\varepsilon} \sum_{(n_x,n_y), (n_x',n_y')} \frac{\varepsilon D_{yx}(\varepsilon) f(E_n) \bigl[ E_{n_y} - E_{n_y'} \bigr] }{\bigl[ \varepsilon^2 + (E_{n_x} - E_{n_x'})^2 \bigr] \bigl[ \varepsilon^2 + (E_{n_y} - E_{n_y'})^2 \bigr]} p_y^{n_yn_y'} p_x^{n_x n_x'}\mathrm{Im}\, j_y^{n'n}(\mathbf{r}),
\end{align}

\section{Calculation of edge current for the semi-infinite 2DEG}

Eigenfunctions of the semi-infinite 2DEG which occupies half-plane $x > 0$ can be parameterized by the wave vector, and they are given as $\psi_{\mathbf{k}} = \sqrt{2/\pi} \sin (k_x x) e^{ik_y y}$. From the equation~\eqref{eq:paramagnetic_simplified} we can see that the diagonal element of the photonic Green's function does not have any contribution to the current, since $\braket{\mathbf{k} | \hat{p}_y | \mathbf{k} '} = \delta ( \mathbf{k} - \mathbf{k} ) k_y$, and $\delta$-function leads to the trivial result.  $\braket{k_x | \hat{p}_x | k_x'} = -i \braket{k_x | \partial_x U | k_x'} / (E_{\mathbf{k}'} - E_{\mathbf{k}})$, and using parity of the photonic Green's function, we obtain
\begin{align}
    j^{\mathrm{d}}_y(\mathbf{r}) = \sum_{\mathbf{k}, \mathbf{k}'} \delta(k_y - k_y') \sum_{\varepsilon} \frac{\varepsilon D_{yx}(\varepsilon) \braket{k_x' | \partial_x U | k_x}}{\varepsilon^2 + (E_{\mathbf{k}'} - E_{\mathbf{k}})^2} \frac{f_{\mathbf{k}'} - f_{\mathbf{k}}}{E_{\mathbf{k}'} - E_{\mathbf{k}}} \psi_{k_x'}(x) \psi_{k_x}^*(x),
\end{align}
Now we want to calculate the integral edge current created $J_{\mathrm{edge}} = \int_0^{\infty} dx j^{\mathrm{d}}_y(x)$. Integration over $x$ results in $\delta(k_x - k_x')$, and therefore, we have
\begin{align}
    J_{\mathrm{edge}} = \sum_{\mathbf{k}, \varepsilon}\frac{\varepsilon D_{yx}(\varepsilon)}{\varepsilon^2 + (E_{\mathbf{k}'} - E_{\mathbf{k}})^2} \braket{k_x | \partial_x U | k_x} \left. \frac{\partial f}{\partial E} \right|_{E = E_ \mathbf{k}}.
\end{align}
Observing now that $\braket{k_x | \partial_x U | k_x} = 2k_x^2 / m = 2k_x \partial E_{\mathbf{k}} / \partial k_x$, we can perform summation over $\mathbf{k}$ and arrive to
\begin{align}
    J_{\mathrm{edge}}  = - \frac{\lambda^2 n_0}{m^2} \sum_\varepsilon \frac{D_{yx}(\varepsilon)}{\varepsilon},
\end{align}
where $n_0$ is density of electrons in the bulk. Summation over $\varepsilon$ results in the expression~\eqref{eq:one_wall_edge}. We can see that this result coincides with the universal result from Eq.~\eqref{eq:Jedg} in the main text.

The paramagnetic contribution is locally zero. Indeed, terms proportional to the diagonal part of the photonic propagator contains odd powers of $k_y$, and thus give zero after integration over this quantum number. Terms with the non-diagonal photonic Green's function matrix elements are also zero which follows from the expression~\eqref{eq:paramagnetic_current}.  

\section{Summation over bosonic Matzubara frequencies for the parabolic confining potential}

We have shown that currents and magnetization for the parabolic potential are proportional to the following summation over Matzubara frequencies that can be calculated explicitly,
\begin{align}
    &\frac{1}{\beta} \sum_{\varepsilon} \frac{\varepsilon D_{yx}(\varepsilon)}{\varepsilon^2 + \omega^2} = \frac{\Delta}{2} \left[ \frac{\Omega_+ [1 + 2n_{\mathrm{B}}(\Omega_+)]}{(\Omega_+^2 - \Omega_-^2)(\omega^2 - \Omega_+^2)} - \frac{\Omega_- [1 + 2 n_{\mathrm{B}}(\Omega_-)]}{(\Omega_+^2 - \Omega_-^2)(\omega^2 - \Omega_-^2)} - \frac{\omega [1 + 2n_{\mathrm{B}}(\omega)]}{(\Omega_+^2 - \omega^2)(\Omega_-^2 - \omega^2)} \right],\\
    &\begin{aligned}
        \frac{1}{\beta} \sum_{\varepsilon} \frac{ \varepsilon D_{yx}(\varepsilon) }{(\varepsilon^2 + \omega^2)^2} = \frac{\Delta}{2} &\left[ -\frac{[1 + 2n_{\mathrm{B}}(\omega) + \omega \partial_\omega n_{\mathrm{B}}(\omega) ]}{2\omega(\omega^2 - \Omega_-^2)(\omega^2 - \Omega_+^2)} + \frac{(\Omega_-^2 \Omega_+^2 - \omega^4)[1 + 2 n_B(\Omega)]}{\omega (\omega^2 - \Omega_-^2)^2 (\Omega^2 - \Omega_+^2)^2} + \right. \\
        &\left.  + \frac{\Omega_-[1 + 2n_{\mathrm{B}}(\Omega_-)]}{(\omega^2 - \Omega_-^2)^2 (\Omega_-^2 - \Omega_+^2)} - \frac{\Omega_+[1 + 2n_{\mathrm{B}}(\Omega_+)]}{(\omega^2 - \Omega_+^2)^2 (\Omega_-^2 - \Omega_+^2)} \right].
    \end{aligned}
\end{align}

\end{widetext}

\bibliography{references}

\end{document}